\documentclass[a4paper,twocolumn]{revtex4}
\usepackage{graphicx}

\begin{document}

\title{Cooperative equilibria in the finite iterated prisoner's dilemma}

\author{Kae Nemoto}

\affiliation{National Institute of Informatics, 2-1-2
Hitotsubashi, Chiyoda-ku, Tokyo 101-0843, Japan}

\author{Michael J. Gagen}

\affiliation{ARC Special Research Centre for Functional and
Applied Genomics, Institute for Molecular Bioscience, University
of Queensland, Brisbane, Qld 4072, Australia}

\email{m.gagen@imb.uq.edu.au}

\date{\today}

\begin{abstract}
Nash equilibria are defined using uncorrelated behavioural or
mixed joint probability distributions effectively assuming that
players of bounded rationality must discard information to locate
equilibria. We propose instead that rational players will use all
the information available in correlated distributions to constrain
payoff function topologies and gradients to generate novel
``constrained" equilibria, each one a backwards induction pathway
optimizing payoffs in the constrained space. In the finite
iterated prisoner's dilemma, we locate constrained equilibria
maximizing payoffs via cooperation additional to the unconstrained
(Nash) equilibrium maximizing payoffs via defection. Our approach
clarifies the usual assumptions hidden in backwards induction.
\end{abstract}


\maketitle

\section{Introduction}

Payoff maximization in the single stage Prisoner's Dilemma (PD)
locates a unique Nash equilibrium point, mutual defection, which
garners players a non-Pareto efficient outcome. Finite repetition
of this single stage game defines the finite Iterated Prisoner's
Dilemma (IPD) which is apparently solved by inductively
propagating the single stage Nash equilibrium of mutual defection
backwards through every stage of the game to establish ``All
Defect" as the unique Nash equilibrium path. This single rational
play strategy prevents players from cooperating to achieve higher
payoffs. Nevertheless, in experimental tests (see
\cite{Cooper_96_18,Milinski_98_13,Davis_99_89,Croson_00_29}),
people often cooperate to garner a greater payoff indicating
either that modelling in game theory is somehow incomplete or that
people behave irrationally. Many different proposals have been
made along these two lines including suggestions to modify
definitions of rationality and to bound rationality
\cite{Radner_80_13,Radner_86_38,Vegaredondo_94_18,Harborne_97_13,Anthonisen_99_14},
to take account of incomplete information
\cite{Harsanyi_67_15,Kreps_82_24,Fudenberg_86_53,Sarin_99_10} and
uncertainty in the number of repeat stages \cite{Neyman_99_45}, to
bound the complexity of implementable strategies
\cite{Neyman_85_22,Rubinstein_86_83,Cho_99_93}, to account for
communication and coordination costs \cite{Raff_00_10}, to
incorporate reputation and experimentation effects
\cite{Evans_97_11} or secondary utility functions as in
benevolence theory \cite{Selten_78_12} or in moral discussions
\cite{Sheng_94_23}, to include adaptive learning
\cite{Groes_99_12} or fuzzy logic \cite{Song_99_63}, or more
directly, to employ comprehensive constructions of normal form
strategy tables \cite{Howard_71,Rapoport_67,Straffin_93}.
Interestingly, quantum correlations can be introduced to resolve
the prisoner's dilemma \cite{Eisert_99_30}.

As noted above, these proposals generally modify the definition of
either the IPD or of player rationality to explain deviations from
the single unique Nash equilibrium pathway. By contrast, in this
paper we assume Common Knowledge of Rationality (CKR) for all
players and no modification of the IPD game definition. Then we
generalize the analysis of the IPD to regimes where the fixed
point theorems underlying existing equilibria do not apply. In
providing an existence proof for mixed strategy equilibria, Nash
and Kuhn made the overly restrictive assumption that each player's
mixed or behavioural strategy choice probability distributions
were continuous and uncorrelated and so strictly independent
\cite{Nash_51_28,Kuhn_1953}. This assumption allowed the analytic
continuation of game payoff functions over a convex probability
polytope enabling the use of fixed point theorems to locate mixed
strategy equilibria \cite{Nash_51_28}. The subsequent widespread
use of this existence proof as a definition of ``Nash equilibria"
requires rational players to locate equilibria by discarding
information inherent in correlations, in effect, an assumption of
``bounded" rationality. For multi-stage games, this same
assumption that players of bounded rationality (employing myopic
agents)3 must discard information by adopting uncorrelated play
allowed Kuhn to define subgame equilibria and subgame
decompositions \cite{Kuhn_1953}. While this assumption is always
valid for single stage games with an empty history set, in
multistage games the joint strategy choice probability
distributions of all players can be more generally correlated
through being conditioned on game history sets, with these
correlations invalidating the {\it a priori} assumption of
uncorrelated mixed or behavioural strategies. Hence, we propose
that players of unbounded rationality, in contrast to those of
``bounded" rationality, will make use of all available information
by exploiting correlated mixed or behavioural joint strategy
choice probability distributions to optimize payoffs.

In this paper, by considering broader classes of correlated joint
probability distributions, we significantly generalize the
analysis of the IPD at the cost of a greater analytic complexity
stemming for instance, from the nonapplicability of fixed point
theorems. It may be questioned whether this cost is a price worth
paying. However, we have seen little in the literature considering
regimes where these {\it a priori} assumptions are invalid, though
these regimes might provide the mathematical arena to model
broader classes of experimental game behaviours. Especially so
given that well established techniques for manipulating
conditioned and correlated joint probability distributions exist.

Previous efforts to model correlated strategy choices have
attempted this task discursively---see for instance the
descriptive derivations of strategy function equilibria in
differential games \cite{Dockner_2000} and trigger function
equilibria in supergames where players encourage adherence to a
cooperation pathway using deviations to ``trigger" credible
punishments
\cite{Friedman_71_1,Friedman_74_22,Friedman_90_31,Friedman_1991,Halpern_01_42}.
These previous treatments have strongly insisted that players
prespecify an outcome for absolutely every possible (and
impossible) situation that might arise in a supergame---for
justifying quotations, see note
\cite{Strat_Quotes}\nocite{Osborne_1995,Fudenberg_2000,Dockner_2000,Friedman_1991}.
As a result, current supergame analysis must either assume that
all mixed or behavioural strategy choice variables are independent
and optimized via a Nash procedure, or that all variables are
fully specified by strategy functions which are themselves
independently selected and again optimized via a Nash procedure.

This paper fills in the missing middle ground here, and considers
correlated mixed or behavioural joint probability distributions
conditioned by sets of strategy functions which specify only some
fraction of the possible variables (defining a dependent set)
while leaving the remaining variables unspecified (an independent
set) to be optimized by standard Nash techniques. (Here, the
dependent variables are described by correlated probability
distributions conditioned on game history sets.) As such, we
introduce no additional properties to the IPD and our use of
strategy functions and Nash optimization techniques are entirely
standard. Optimizing functions over a set of independent variables
given a set of constrained dependent variables is the subject
matter of constrained optimization theory as in Lagrange
multipliers \cite{Stewart_1999,Anton_1995}, variational calculus
\cite{Weinstock_52,Gelfand_63,Clegg_68}, and dynamic programming
\cite{Bellman_57}. Each adopted constraint set generalizes the
{\it a priori} assumption made by Nash and Kuhn that each player's
mixed or behavioural strategy choice probability distributions are
uncorrelated and separable. In fact, Nash and Kuhn's multistage
analysis corresponds to the special case where the constraint set
is empty, and Nash equilibria are defined only for empty
constraint sets. Hence, in our generalized analysis (as in any
constrained optimization procedure), we must first take account of
applied constraint sets prior to locating ``constrained
equilibria" in the constrained probability space. (We note that
games with mixed strategies constrained to lie within convex
hyperpolyhedron by linear inequalities and equations have been
considered in Refs. \cite{Vajda_1966,Owen_1995}.)

To avoid any confusion about our generalization, we first review
the original Nash equilibrium definitions in the next section
(\ref{sect_review_Nash}), and then generalize these definitions to
define constrained equilibria applicable to non-empty constraint
sets in the section following (\ref{sect_constrained_eq}).  In
section \ref{sect_IPD}, we derive constrained equilibria in the
IPD, and observe the different player behaviours arising from
adopted constraints. Further, we apply the Nash equilibria
definitions on the constrained equilibrium space to specify global
equilibria. Finally we discuss the role of the backwards induction
argument as it is applied to finite supergames in section
\ref{sect_BI}.

\section{Review of Nash equilibria}
\label{sect_review_Nash}

Throughout this paper, we consider supergames formed by finitely
iterating a single stage game over $1\leq n\leq N$ stages where
each stage is played between two non-communicating rational
players denoted $P_x$ and $P_y$. In the $n^{\rm th}$ stage,
players $P_x$ and $P_y$ choose their respective stage strategies,
denoted $x_n$ and $y_n$, from the same strategy set $S$, that is
$x_n,y_n\in S$, where $S = \{ s_1 , \dots, s_s \}$ and $s$ is the
total number of strategies, or alternatively, $S=\{ s |
s\in\mbox{R}, s_1\leq s\leq s_s\}$ when strategy choice is
continuous.  Game history sets $H_n$ known to both players in
stage $(n+1)$ record occurring events with $H_0=\emptyset$, and
$H_n=\{x_1,\dots,x_n,y_1,\dots,y_n\}$. The payoffs $\Pi_x$ and
$\Pi_y$ for players $P_x$ and $P_y$ are each specified as mappings
(or functions) from the set of all chosen strategies
$H_N=\{x_1,\dots,x_N,y_1,\dots,y_N\}$ to the real line
\cite{Hart_92_c2,Sorin_92_c4} via
\begin{equation}
  \Pi_z = \Pi_z (x_1,\dots,x_N,y_1,\dots,y_N), \;\;\; z\in\{x,y\}.
\end{equation}
We consider supergames where these mappings are defined as
summations of the respective $n^{\rm th}$ stage player payoffs
$\pi_x (x_n,y_n)\geq 0$ and $\pi_y (x_n,y_n)\geq 0$ assumed
non-negative without loss of generality, giving
\begin{equation}
 \Pi_z  = \sum_{n=1}^{N} \pi_z (x_n,y_n), \;\;\; z\in\{x,y\},
\end{equation}
nominally functions of $2N$ variables $\{ x_1,\dots , x_N\}$ and
$\{ y_1,\dots, y_N\}$. The goal of each player is to maximize
their respective total payoff functions $\Pi_x$ and $\Pi_y$.

Strategy choices are defined as a functional mapping from the game
history sets $H_n$ to the strategy set $S$, which specify after
each history the specific strategy choices $x_n$ and $y_n$ to be
selected \cite{Hart_92_c2,Sorin_92_c4}. Thus, following
standard definitions \cite{Hart_92_c2}:
\begin{quote}
 A pure strategy [$x_n$] of player [$P_x$] is a function
 \begin{equation}     \label{eq_dep_var}
  x_n:H_{n-1} \rightarrow S=\{s_1,s_2,\dots\}.
 \end{equation}
\end{quote}

In Ref. \cite{Nash_51_28}, Nash defined mixed strategy equilibria
in terms of expected value payoff functions.  The most general
possible expected payoff functions are
\begin{eqnarray}     \label{eq_expected_payoff1}
\langle \Pi_z \rangle &=&
 \sum_{x_1\dots x_N, y_1\dots y_N=s_1}^{s_s}
 P(x_1,\dots,x_N,y_1,\dots,y_N)  \nonumber \\
 & & \times \Pi_z(x_1,\dots,x_N,y_1,\dots,y_N)
\end{eqnarray}
for $z\in\{x,y\}$, and where the probability that player $P_x$
($P_y$) chooses strategy choice $x_n\in S$ ($y_n\in S$) in stage
$n$ for $1\leq n\leq N$ is $P(x_1,\dots,x_N,y_1,\dots,y_N)$. Here,
the average is calculated over an ensemble of trials representing
every possible circumstance and outcome. In particular, the total
ensemble consists of an infinite number of sub-ensembles, one for
every possible joint probability distribution, each one of which
contains an infinite number of trial outcomes. This is in accord
with von Neumann and Morgenstern's definition of a strategy as ``a
complete plan: a plan which specifies what choices [a player] will
make in every possible situation, for every possible actual
information which [that player] may possess at that moment"
\cite{vonNeumann_44}. We emphasize that a player's complete
strategy list is not synonymous with a mere listing of all the
possible choices determining every pathway through the complete
game tree despite this common usage in Ref.
\cite{vonNeumann_44}---such a listing is missing information.  A
full listing of every possible situation which might occur will
contain a successive listing of all the many possible joint
strategy choice probability distributions which might be adopted
by the players as well as subsidiary information about all the
possible pathways through the associated game tree generated under
those adopted distributions. If the information about the joint
probability distributions is absent, this is equivalent to making
a default assumption that all pathways are equally weighted (as
conversely, differing joint probability distributions weight
pathways differently).  In actuality, discarding information about
adopted joint probability distributions is equivalent to an
assumption of bounded rationality, and the necessity of such
restrictions has never been demonstrated.  In particular, von
Neumann and Morgenstern specifically asserted that they were using
a method of ``indirect proof" to imagine the form of a successful
theory and to test the consequences for problems and
contradictions \cite{vonNeumann_44}. Naturally, such
contradictions were never found as restricting the solution space
to a valid sub-ensemble ensures the absence of contradictions
though at the expense of incomplete results. In this paper we do
not make this unjustified assumption. As usual then, using $P(A
\mbox{ and }B)=P(A)P(B|A)$ we have
\begin{eqnarray}
  P(x_1,\dots,x_N,y_1,\dots,y_N)&=&  P_{1x}(x_1) P_{1y}(y_1) \times \nonumber \\
  &&   \hspace{-1cm} \times P(x_2,\dots,y_2,\dots|H_1),
  \label{gen_prob1}
\end{eqnarray}
and the iterated identity
\begin{eqnarray}
  && \hspace{-.5cm} P(x_n,\dots,x_N,y_n,\dots,y_N|H_{n-1})=P_{nx}(x_n|H_{n-1}) \times \nonumber \\
  &&   P_{ny}(y_n|H_{n-1})
     P(x_{n+1},\dots,y_{n+1},\dots|H_n),
  \label{gen_prob}
\end{eqnarray}
successively applied for all $n$. Here, same stage choices $x_n$
and $y_n$ are independent events conditioned on the history set
$H_{n-1}$ and so potentially correlated. We also elect here to
maintain the time ordering of conditioning events, though this is
not necessary---probability distributions can be pre- or
post-conditioned, so two events $A$ and $B$ occurring at two
different times have joint probability $P(A\mbox{ and
}B)=P(A)P(B|A)=P(B)P(A|B)$.  So also, a player in their pregame
analysis can condition events in one stage $n$ on either earlier
or later stage events as desired. The most general possible
expected payoff functions are then
\begin{eqnarray}    \label{eq_general_payoff}
\langle \Pi_z \rangle &=&
    \sum_{\stackrel{s_1\leq x_n,y_n\leq s_s}{H_{n-1}}}
     P(H_{n-1}) P(x_n,y_n|H_{n-1})\Pi_z(x_1,\dots) \nonumber \\
 &=& \sum_{\stackrel{x_1\dots x_N}{y_1\dots y_N}}
 P_{1x}(x_1) P_{1y}(y_1) \times \dots \times P_{Nx}(x_N|H_{N-1}) \nonumber \\
 &&  \times  P_{Ny}(y_N|H_{N-1})
   \Pi_z(x_1,\dots,x_N,y_1,\dots,y_N)  \nonumber \\
 &=& \sum_{n=1}^{N} \sum_{\stackrel{x_1\dots x_n}{y_1\dots y_n}}
 P_{1x}(x_1) P_{1y}(y_1) \times \dots  \nonumber \\
 &&  \times P_{nx}(x_n|H_{n-1}) P_{ny}(y_n|H_{n-1}) \pi_z(x_n,y_n),
\end{eqnarray}
for $z\in\{x,y\}$. Each of the conditioned distributions
$P_{nz}(z_n|H_{n-1})$ can be written as a list of potentially
correlated behavioural strategy distributions
\begin{equation}
   P_{nz}(z_n|H_{n-1}) = \left\{
    \begin{array}{cc}
      P_{nz,H'_{n-1}}(z_n), & \mbox{if } H_{n-1}=H'_{n-1} \\
       &  \\
      P_{nz,H''_{n-1}}(z_n), & \mbox{if } H_{n-1}=H''_{n-1} \\
       &  \\
      \vdots & \vdots ,\\
    \end{array}
   \right.
\end{equation}
with up to $2^{2(n-1)}$ entries, one for each possible history set
$H_{n-1}$.  The individual distributions $P_{nz,H_{n-1}}(z_n)$ and
$P_{n'z',H_{n'-1}}(z'_{n'})$ can still be correlated as when, for
instance, a single ``dice" is used to determine both outcomes.
(For completeness, note \cite{Correlations} lists the definitions
of correlated variables in terms of their covariance, variance and
means.) Such correlations further imply  that these distributions
are not necessarily continuous.  These potentially correlated
behavioural strategy distributions allow writing the most general
expected payoff functions as
\begin{eqnarray}
 \langle \Pi_z \rangle &=&
 \sum_{\stackrel{x_1\dots x_N}{y_1\dots y_N}}
 P_{1x}(x_1) P_{1y}(y_1) \times \dots \times P_{Nx,H_{N-1}}(x_N) \nonumber \\
 &&  \times  P_{Ny,H_{N-1}}(y_N)
   \Pi_z(x_1,\dots,x_N,y_1,\dots,y_N).
\end{eqnarray}
Here, all possible contingent histories have been taken into
account weighted by their respective conditioned probabilities.
Players can now seek to optimize their payoffs by applying any
relevant optimization technique to these general expected payoff
functions. Of course, if the functions are correlated and
discontinuous then fixed point theorems cannot be used to locate
optima, and also, if the variables are correlated it is absolutely
necessary to resolve the correlations as imposed constraints prior
to applying any optimization procedure.  That is, if correlations
exist, any optimization procedure such as backwards induction must
take those correlations into account as imposed constraints before
seeking to derive an optimal pathway.

Given these most general expected value payoff functions, we now
revisit the definition of Nash equilibria as developed by Nash
\cite{Nash_51_28} and as explicated by Kuhn \cite{Kuhn_1953}.
When introducing behavioral strategies Kuhn ``explicitly assumed
that the choices of alternatives at different history sets are
made independently. Thus it might be reasonable to call them
`uncorrelated' or `locally randomized' strategies."
\cite{Kuhn_1953}.  Such uncorrelated behavioral strategies capture
the myopic viewpoint of non-communicating agents possessing ``a
local perspective [which] decentralizes the strategy decision of
player $i$ into a number of local decisions."
\cite{vanDamme_92_41}.  In this, the agent-normal game form,
myopic agents at each history set determine paths through the game
tree using probability distributions which are uncorrelated and
independent.  This assumption allowed Kuhn to prove the
equivalence of uncorrelated behavioural strategies and the
uncorrelated mixed strategies introduced by Nash in games of
perfect recall \cite{Kuhn_1953}. This equivalence was established
by recognizing that player $P_x$ ($P_y$) could index all their
possible pure strategies by parameter $\alpha$ ($\beta$) with the
probability of playing that strategy being $P_x(\alpha)$
($P_y(\beta)$) given by an appropriate product of the uncorrelated
behavioural stategies $P_{nx,H_{n-1}}(x_n)$
($P_{ny,H_{n-1}}(y_n)$).  This then allowed writing the
non-general expected payoff functions as
\begin{equation}    \label{eq_general_payoff_Nash}
\langle \Pi_z \rangle =
 \sum_{\alpha \beta} P_{x}(\alpha) P_{y}(\beta)  \Pi_z(\alpha,\beta)
\end{equation}
for $z\in\{x,y\}$ where here, the summation is over an
appropriately limited set of $\alpha$ and $\beta$ values. Nash
considered the mixed strategies $P_x(\alpha)$ and $P_y(\beta)$ as
``a collection of non-negative numbers which have unit sum and are
in one to one correspondence with his pure strategies." so the
expected payoff functions were linear in the mixed strategies for
each player allowing optimization over a ``convex subset of a real
vector space" via fixed point theorems \cite{Nash_51_28}. This
definition follows that of von Neumann and Morgenstern
\cite{vonNeumann_44} in establishing a one to one correspondence
between a player's pure and mixed strategies which are subject to
appropriate normalization constraints but to no other constraints
such as might result from using fully general joint strategy
choice probability distributions. Of course, these restrictive
assumptions limit the ensemble over which payoff averages are
calculated, and in the full ensemble correlated behavioural
strategies can break the one-to-one correspondence between the
mixed strategies and the full set of uncorrelated pure strategies,
can render expected payoff functions discontinuous, and can
invalidate the use of fixed point theorems as an optimization
technique.

The assumption that players employ uncorrelated mixed or
behavioural strategy choices then allows the definition of Nash
equilibria in terms of the probabilities $\bar{p}\equiv
\{P_x(\alpha),\forall \alpha\}$ and $\bar{q}\equiv
\{P_y(\beta),\forall \beta\}$. Following Nash \cite{Nash_51_28},
\begin{quote}
  {\bf If and only if} each player's mixed or behavioural strategy choices are
  uncorrelated and independent, then a
  $2$-tuple $(\bar{p}^*,\bar{q}^*)$ of unconditioned probability
  distributions forms a mixed strategy equilibrium point
  if and only if for all players,
  \begin{eqnarray} \label{Nash_mixed}
     \langle \Pi_x (\bar{p}^*,\bar{q}^*)\rangle &=&
        \max_{\forall \bar{p}}
        [\langle \Pi_x(\bar{p},\bar{q}^*)\rangle] \nonumber \\
     \langle \Pi_y (\bar{p}^*,\bar{q}^*)\rangle &=&
        \max_{\forall \bar{q}}
        [\langle \Pi_y(\bar{p}^*,\bar{q})\rangle].
  \end{eqnarray}
  Thus an equilibrium point is a $2$-tuple $(\bar{p}^*,\bar{q}^*)$
  such that each player's pure strategy maximizes their payoff if
  the strategies of the others are held fixed.
  Thus each player's strategy is  optimal against those of the
  others \cite{Nash_51_28}.
\end{quote}
We note here that Nash proved an existence theorem in Ref.
\cite{Nash_51_28} and not a uniqueness theorem, and it is well
known that when the restrictive assumptions are not made then
mixed strategy equilibria do not necessarily exist
\cite{Osborne_1995,Fudenberg_2000,Sion_57,dAspremont_79_11,Dasgupta_86_1,Dasgupta_86_27,Aumann_74_67}.

Pure strategy Nash equilibria can be defined by further
specializing the strategy choice probability distributions to be
either zero or one, $P_x(\alpha),P_y(\beta)\in\{0,1\}$ for all
values of $\alpha$ and $\beta$, so one or another pure strategy is
independently chosen by each player with certainty. Then pure
strategy Nash equilibria can be defined following Nash
\cite{Nash_51_28}:
\begin{quote}
  {\bf If and only if} each player's pure strategy choices are
  uncorrelated and independent, then a
  $2$-tuple $(\alpha^*,\beta^*)$ is a pure strategy
  equilibrium point if and only if for all players,
  \begin{eqnarray} \label{Nash_pure}
     \Pi_x (\alpha^*,\beta^*) &=&
        \max_{\forall \alpha} [\Pi_x(\alpha,\beta^*)] \nonumber \\
     \Pi_y (\alpha^*,\beta^*) &=&
        \max_{\forall \beta} [\Pi_y(\alpha^*,\beta)].
  \end{eqnarray}
\end{quote}

Another situation where the Nash definition can be applied is when
all variables are fully specified by strategy functions which are
themselves independently selected.  In this case, strategy choices
are conditioned on earlier events and so possibly correlated in
any stage despite being chosen independently by each player
\cite{Dockner_2000,Friedman_71_1,Friedman_74_22,Friedman_90_31,Friedman_1991}.
However strategy functions introduced in these approaches are very
limited in the sense that any implemented strategy function set
must fully specify an outcome for every stage and every possible
situation which might arise in a game. (For justifying quotations
see \cite{Strat_Quotes}.) Because players $P_x$ and $P_y$
independently choose sets of $N$ strategy functions denoted ${\cal
A}_x=\{x_n:H_{n-1}\rightarrow S, 1\leq n\leq N\}$ and similarly
for ${\cal A}_y$ to generate payoffs $\Pi_z({\cal A}_x,{\cal
A}_y)$ for $z=\{x,y\}$, it is possible to define strategy function
Nash equilibria \cite{Dockner_2000,Nash_51_28}:
\begin{quote}
  {\bf If and only if} all $2N$ strategy choice variables are
  functionally specified,
  a $2$-tuple strategy profile $\phi=\{{\cal A}_x^*,{\cal A}_y^*\}$
  is a strategy function Nash equilibria
  if and only if for all players,
  \begin{eqnarray}
     \Pi_x ({\cal A}_x^*,{\cal A}_y^*) &=&
        \max_{\forall {\cal A}_x} [\Pi_x({\cal A}_x,{\cal A}_y^*)] \nonumber \\
     \Pi_y ({\cal A}_x^*,{\cal A}_y^*) &=&
        \max_{\forall {\cal A}_y} [\Pi_y({\cal A}_x^*,{\cal A}_y)].
  \end{eqnarray}
\end{quote}
Essentially equivalent definitions underly trigger function
equilibria
\cite{Friedman_71_1,Friedman_74_22,Friedman_90_31,Friedman_1991,Dockner_2000}.

Based on the equilibria definitions above, supergame analysis must
either assume that all variables are independent and optimized via
a Nash procedure, or that all variables are fully specified by
strategy functions.  To treat more general strategy
functions, it is necessary to extend the Nash equilibrium concept
to regimes where fixed point theorems are inapplicable. The next
section does this by defining new constrained equilibria.

\section{Constrained equilibria}
\label{sect_constrained_eq}

In this section we define constrained equilibria through allowing
the use of correlated multivariate probability distributions to
optimize expected payoffs.

We note firstly that neither Nash nor Kuhn provided any rationale
requiring rational players to restrict the size of the ensemble
used to calculate payoff averages by only employing uncorrelated
mixed or behavioural probability distributions. Likely, this
assumption was made in the context of an existence proof to
simplify analysis as the fully general treatment of correlated
multivariate probability distributions is difficult. However,
adopting correlated distributions can often greatly simplify
analysis. Consider that when events $A$ and $B$ are perfectly
correlated, the joint probability of both events reduces to
$P(A\mbox{ and }B)=P(A)P(B|A)=P(A)$, so one variable entirely
disappears reducing the dimensionality of the problem space and
simplifying the problem.  (This dimensionality reduction is a
normal result whenever constraints are applied in optimization
problems.)

In fact, any assumption that players must adopt the restricted
ensemble available under uncorrelated play is equivalent to a
claim that players must discard information and so amounts to an
assumption of bounded rationality.  (For completeness, note
\cite{Information} details the mutual information content of
correlated joint probability distributions.) How might rational
players exploit the information in correlated distributions?  A
minimum necessary condition for the existence of an equilibrium
point is that opponents must be unable to improve their payoffs by
altering their strategy choices, so essentially, opponent's payoff
function ``gradients" must be negative at equilibrium points. Now,
rational players can exploit correlations to constrain the payoff
function space topology so as to alter the possible directions in
which payoff functions can change. As ``gradients" can only be
taken along allowed directions, such constraints alter the payoff
function ``gradients" at any point. Thus, it is possible for
rational players to choose correlations to alter payoff function
``gradients"  to generate novel equilibria at novel points. In
this paper, we assume that rational players will make use of all
available information including that implicit in correlated joint
probability distributions.

In this paper, constrained equilibria are defined for pairs of
sets of strategy functions $\{\{X_n\},\{Y_n\}\}$. For the pure
strategy case, these strategy functions $Z_n$ may be represented
as
\begin{eqnarray} \label{strategy_functions}
  x_n &=& X_n (x_1,\dots,x_{n-1},y_1,\dots,y_{n-1}) \nonumber \\
  y_n &=& Y_n (x_1,\dots,x_{n-1},y_1,\dots,y_{n-1}).
\end{eqnarray}
Here $x_n$ and $y_n$ might be independent variables or dependent
on some or all of the variables
$x_1,\dots,x_{n-1},y_1,\dots,y_{n-1}$.  For the pure strategy
case, neither these variables nor the strategy functions are
probabilistic.  In terms of conditioned probability distributions,
these constraints take the form
\begin{eqnarray} \label{strategy_prob_dist}
  P_{nx}(x_n|H_{n-1}) &=& \delta_{x_n,X_n(H_{n-1})} \nonumber \\
  P_{ny}(y_n|H_{n-1}) &=& \delta_{y_n,Y_n(H_{n-1})},
\end{eqnarray}
where $\delta_{a,b}$ is one if $a=b$ and zero otherwise.  (Of
course, more general distributions could be considered.) These
functional notations, though widely used to represent payoff
functions, have not been widely employed for strategy functions.
We note that it has been used in the derivation of best reply
(strategy) functions \cite{Friedman_1991,Fudenberg_2000},
Stackelberg duopolies \cite{Fudenberg_2000}, a two-stage
prisoner's dilemma \cite{Gibbons_1992}, correlation in randomized
strategies \cite{Aumann_74_67}, and differential games
\cite{Dockner_2000}, while a number of texts describe strategies
as ``functions" without actually introducing a function notation
\cite{Hart_92_c2,Sorin_92_c4,Owen_1995,Osborne_1995}.

Suppose player $P_x$ (or $P_y$) chooses a set of strategy
functions, which we might conveniently call their algorithm
denoted as ${\cal A}_x=\{X_m,\dots,X_k\}$ (or ${\cal
A}_y=\{Y_n,\dots,Y_j\}$), to constrain some (or none) of their
strategy choice variables creating a set of dependent variables
and a remaining set of independent variables. For notational
convenience, we relabel the independent variable sets for $P_x$
and $P_y$ as respectively
$\bar{\alpha}=\{u_1,\dots,u_{\hat{x}}\}\in S^{\times \hat{x}}$ and
$\bar{\beta}=\{v_1,\dots,v_{\hat{y}}\}\in S^{\times \hat{y}}$
where $0<\hat{x}+\hat{y}\leq 2N$. We assume here that at least one
variable remains independent as otherwise, the optimization
becomes trivial. Immediately then, the payoff functions for the
players become composite with reduced dimensionality (and changed
properties) given by
\begin{equation} \label{new_payoff}
   \Pi_z \to \Pi_z (\bar{\alpha},\bar{\beta}),   \;\;\;\; z\in\{x,y\}.
\end{equation}
Here, no dependent variables, such as $x_2=X_2(x_1,y_1)$ say,
appear in the composite payoff functions.

With the representation of strategy functions in terms of the
independent variables, the Nash equilibrium definition can be
applied to the space of independent variables to define pure
strategy constrained equilibria via:
\begin{quote}
   Given a particular constraint set ${\cal A}_x$ and ${\cal A}_y$,
   a $2$-tuple $(\bar{\alpha}^*,\bar{\beta}^*)$ is a pure strategy
   constrained equilibria if and only if for all players,
  \begin{eqnarray}
     \Pi_x (\bar{\alpha}^*,\bar{\beta}^*) &=&
        \max_{\forall \bar{\alpha}} [\Pi_x(\bar{\alpha},\bar{\beta}^*)] \nonumber \\
     \Pi_y (\bar{\alpha}^*,\bar{\beta}^*) &=&
        \max_{\forall \bar{\beta}} [\Pi_y(\bar{\alpha}^*,\bar{\beta})].
  \end{eqnarray}
\end{quote}

We now generalize this definition to mixed strategy constrained
equilibria. In this case, strategies at any stage can be
probabilistic, so the strategy functions of Eqs.
(\ref{strategy_functions}) and (\ref{strategy_prob_dist}) applied
at stage $n$ map each history set $H_{n-1}$ to a probability
distribution.  The set of these probabilistic strategy functions
$P_{nz}(z_n|H_{n-1})$ forms an algorithm ${\cal A}_z$. For a pair
of algorithms $\{ {\cal A}_x,{\cal A}_y\}$, the expected total
payoffs given as Eq. (\ref{eq_expected_payoff1}) can be rewritten
in terms of independent probability distributions $\{
\bar{p},\bar{q}\}$ only, where $\bar{p}=\{p_1,\ldots\}$ specifies
the probability that any allowed value of $\bar{\alpha}$ is
implemented, while $\bar{q}=\{q_1,\ldots\}$ specifies the
probability that any allowed value of $\bar{\beta}$ is
implemented.  Here we used the same relabelling as in the pure
strategy case above. Then, we can define mixed strategy
constrained equilibria as:
\begin{quote}
  Given a particular constraint set ${\cal A}_x$ and ${\cal A}_y$,
   a $2$-tuple $(\bar{p}^*,\bar{q}^*)$ is a mixed strategy
  constrained equilibria if and only if for all players,
  \begin{eqnarray}
     \langle \Pi_x (\bar{p}^*,\bar{q}^*)\rangle &=&
        \max_{\forall \bar{p}}
        [\langle \Pi_x(\bar{p},\bar{q}^*)\rangle] \nonumber \\
     \langle \Pi_y (\bar{p}^*,\bar{q}^*)\rangle &=&
        \max_{\forall \bar{q}}
        [\langle \Pi_y(\bar{p}^*,\bar{q})\rangle].
  \end{eqnarray}
\end{quote}

As is usual in optimization theory, every alternate strategy
function set $\{{\cal A}_x,{\cal A}_y\}$ imposes constraints on
either the space of possible pure strategy choice variables or the
space of possible mixed strategy probability distributions.
Geometrically, these constraints take a cross-section onto some
subspace wherein all constraints are satisfied, and in which the
composite payoff functions exhibit changed continuity properties
and altered maxima.  The composite payoff functions of reduced
dimensionality define pruned extensive form game trees involving
only independent variables---variational optimization techniques
can only ever be applied to independent variables. Consequently,
subgame decompositions, and Nash equilibria can be applied to
extensive form game trees only after these have been pruned of
dependent variables. As is usual in constrained optimization
problems, different constraint sets generate novel trees defining
novel equilibria. In the next section we demonstrate how these new
equilibria emerge in the IPD analysis. A question which might
arise here is how to determine the best equilibrium among these
new equilibria. As is well known in game theory, in general, there
is no simple way to choose between many alternate Nash equilibria
\cite{Straffin_93,Hargreaves_95,Osborne_1995,Fudenberg_2000}.
However, through the example of an IPD in the next section, we
will show a way to address this issue using standard Nash
techniques.

\section{Constrained equilibria in the finite iterated prisoner's dilemma}
\label{sect_IPD}

In this section we determine constrained equilibria in the finite
IPD and demonstrate that cooperation can naturally emerge as a
consequence of constraints. In this supergame, each player has two
possible single stage strategy choices $S=\{C,D\}$ for Cooperate
and Defect respectively with stage payoffs $\pi_x (x_n,y_n)\geq 0$
and $\pi_y (x_n,y_n)\geq 0$ determined by the payoff matrix
\begin{equation}\label{eq_2pIPD}
  \begin{array}{cc}
      & P_y \\
    P_x &
    \begin{array}{c|cc}
      (\pi_x,\pi_y)  & C & D \\   \hline
      C & (2,2) & (0,3) \\
      D & (3,0) & (1,1). \\
    \end{array}
  \end{array}
\end{equation}
This payoff matrix defines single stage payoff functions
\begin{eqnarray}        \label{eq_pi}
  \pi_x (x_n,y_n) & = &  2 + x_n - 2 y_n    \nonumber \\
  \pi_y (x_n,y_n) & = &  2 - 2 x_n + y_n ,
\end{eqnarray}
where $z_n$ represents the strategy choice for player $P_z$ such
that $0$ represents cooperation and $1$ represents defection.
Total game payoffs of a finite IPD of the length $N$ are then
\begin{eqnarray}   \label{ind}
  \Pi_x (x_1\dots x_N,y_1\dots y_N) &=& \sum_{n=1}^N (2 + x_n - 2 y_n )  \nonumber \\
  \Pi_y (x_1\dots x_N,y_1\dots y_N) &=& \sum_{n=1}^N (2 - 2 x_n + y_n )  .
\end{eqnarray}
Following Eq. \ref{eq_general_payoff}, the expected payoff
functions for players $P_x$ and $P_y$ are then
\begin{eqnarray}    \label{}
\langle \Pi_x \rangle &=& 2N +
  \sum_{n=1}^{N} \sum_{\stackrel{x_1\dots x_n}{y_1\dots y_n}}
 P_{1x}(x_1) P_{1y}(y_1) \times \dots  \nonumber \\
 &&  \times P_{nx,H_{n-1}}(x_n) P_{ny,H_{n-1}}(y_n)
 (x_n - 2 y_n) ,  \nonumber \\
\langle \Pi_y \rangle &=& 2N +
  \sum_{n=1}^{N} \sum_{\stackrel{x_1\dots x_n}{y_1\dots y_n}}
 P_{1x}(x_1) P_{1y}(y_1) \times \dots  \\
 &&  \times P_{nx,H_{n-1}}(x_n) P_{ny,H_{n-1}}(y_n)
 (-2x_n + y_n). \nonumber
 \end{eqnarray}
We now derive the unconstrained Nash equilibrium for the IPD after
applying the assumption of bounded rationality so all probability
distributions are uncorrelated.  We first note that the total rate
of change of the expected payoff function with respect to the
changing probability distributions is
\begin{equation}
  \frac{d \langle \Pi_z \rangle}{d \left[ P_{nz,H_{n-1}}(1)\right]}
   = \frac{\partial \langle \Pi_z \rangle}{\partial \left[ P_{nz,H_{n-1}}(1)\right]}
    - \frac{\partial \langle \Pi_z \rangle}{\partial \left[
    P_{nz,H_{n-1}}(0)\right]}
\end{equation}
due to the normalization constraint
$P_{nz,H_{n-1}}(0)=1-P_{nz,H_{n-1}}(1)$.  The shorthand notation
$H_{n}=\{H_{n-1},x_n,y_n\}$ and some algebra allows writing the
optimization conditions for player $P_x$ as the set of
simultaneous equations
\begin{eqnarray}
 \frac{d \langle \Pi_x \rangle}{d [P_{1x}(1)]}&=& \dots,  \nonumber \\
 &\vdots&   \nonumber \\
 \frac{d \langle \Pi_x \rangle}{d [P_{(N-1)x,H_{N-2}}(1)]}
 &=& 1 +  \nonumber \\
 && \hspace{-3cm} \sum_{\stackrel{x_1\dots x_{N-2}}{y_1\dots y_{N-2}}}
       P_{1x}(x_1) \dots P_{(N-2)y,H_{N-3}}(y_{N-2}) \times \nonumber \\
 && \hspace{-3cm}
    \sum_{y_{N-1}} P_{(N-1)y,H_{N-2}}(y_{N-1}) \sum_{x_N y_N} (x_N-2y_N) \times \nonumber \\
 && \hspace{-3cm} \left[ \right. P_{Nx,\{H_{N-2},1,y_{N-1}\}}(x_N)
        P_{Ny,\{H_{N-2},1,y_{N-1}\}}(y_N) - \nonumber \\
 && \hspace{-3cm} P_{Nx,\{H_{N-2},0,y_{N-1}\}}(x_N)
           P_{Ny,\{H_{N-2},0,y_{N-1}\}}(y_N) \left. \right], \nonumber \\
 \frac{d \langle \Pi_x \rangle}{d [P_{Nx,H_{N-1}}(1)]} &=& 1.
\end{eqnarray}
The equivalent simultaneous optimization conditions for player
$P_y$ are
\begin{eqnarray}
 \frac{d \langle \Pi_y \rangle}{d [P_{1y}(1)]}&=& \dots,  \nonumber \\
 &\vdots&   \nonumber \\
 \frac{d \langle \Pi_y \rangle}{d [P_{(N-1)y,H_{N-2}}(1)]}
 &=& 1 +  \nonumber \\
 && \hspace{-3cm} \sum_{\stackrel{x_1\dots x_{N-2}}{y_1\dots y_{N-2}}}
       P_{1x}(x_1) \dots P_{(N-2)y,H_{N-3}}(y_{N-2}) \times \nonumber \\
 && \hspace{-3cm}
    \sum_{x_{N-1}} P_{(N-1)x,H_{N-2}}(x_{N-1}) \sum_{x_N y_N} (y_N-2x_N) \times \nonumber \\
 && \hspace{-3cm} \left[ \right.
        P_{Nx,\{H_{N-2},x_{N-1,1}\}}(x_N)
        P_{Ny,\{H_{N-2},x_{N-1},1\}}(y_N) - \nonumber \\
 && \hspace{-3cm}
           P_{Nx,\{H_{N-2},x_{N-1},0\}}(x_N)
           P_{Ny,\{H_{N-2},x_{N-1},0\}}(y_N) \left. \right], \nonumber \\
 \frac{d \langle \Pi_y \rangle}{d [P_{Ny,H_{N-1}}(1)]} &=& 1.
\end{eqnarray}
Subsequently each player, denoted $P_z$, solves their respective
sets of simultaneous equations to maximize their payoff by setting
$P_{Nz,H_{N-1}}=1$ for all history sets $H_{N-1}$, and by setting
$P_{(N-1)z,H_{N-2}}=1$ for all history sets $H_{N-2}$, and so on.
The final result is that both players defect at every stage giving
the optima as $(x_n,y_n)=(1,1)\equiv(D,D)$ for all $n$. At this
point, payoffs are
$(\langle\Pi_x\rangle,\langle\Pi_y\rangle)=(N,N)$.

Now, we generate constrained equilibria using correlated
distributions in the most general expected payoff functions of Eq.
(\ref{eq_general_payoff}).  As a first step, we consider player
$P_x$ adopts Markovian-like (MKV) strategy functions dependent
only on the results of the previous stage via
\begin{equation}     \label{eq_Mark_x}
  x_n = X_n(y_{n-1}) = y_{n-1},
\end{equation}
for $2\leq n\leq N$.  We assume $P_y$ adopts an empty constraint
set so all $P_{ny,H_{n-1}}(y_n)$ distributions are independent.
The imposed constraints are equivalent to the correlated
probability distributions
$P_{nx}(x_n|H_{n-1})=\delta_{x_n,y_{n-1}}$, so the most general
expected payoff functions become
\begin{eqnarray}   \label{eq_MKV_IND}
\langle \Pi_z \rangle &=&
   \sum_{x_1,y_1,\dots,y_N}
 P_{1x}(x_1) P_{1y}(y_1)P_{2y}(y_2|H_1) \times \dots \nonumber \\
 &&  \times  P_{Ny}(y_N|H_{N-1})
   \Pi_z(x_1,y_1,\dots,y_N)
\end{eqnarray}
for $z\in\{x,y\}$, with generated payoffs
\begin{eqnarray}
 \Pi_x(x_1,y_1,\dots,y_N) & = & 2N + x_1 - \sum^{N-1}_{n=1} y_n  - 2 y_N  \nonumber \\
 \Pi_y(x_1,y_1,\dots,y_N) & = & 2N - 2 x_1 - \sum^{N-1}_{n=1} y_n  + y_N.
\end{eqnarray}
The expected payoff functions for players $P_x$ and $P_y$ are then
\begin{eqnarray}    \label{}
\langle \Pi_x \rangle &=& 2N + \sum_{x_1} P_{1x}(x_1) x_1 +\nonumber \\
  && -  \sum_{n=1}^{N-1} \sum_{y_1\dots y_n}
  P_{1y}(y_1) \dots  P_{ny,H_{n-1}}(y_n) y_n + \nonumber \\
  &&  -2 \sum_{y_1\dots y_N} P_{1y}(y_1) \dots
  P_{Ny,H_{N-1}}(y_N) y_N,  \nonumber \\
\langle \Pi_y \rangle &=& 2N -2 \sum_{x_1} P_{1x}(x_1) x_1 +  \\
  && - \sum_{n=1}^{N-1} \sum_{y_1\dots y_n}
  P_{1y}(y_1) \dots  P_{ny,H_{n-1}}(y_n) y_n + \nonumber \\
  && \sum_{y_1\dots y_N} P_{1y}(y_1) \dots  P_{Ny,H_{N-1}}(y_N) y_N . \nonumber
\end{eqnarray}
The generated constrained equilibria are now calculated by
applying the assumption of bounded rationality so all remaining
distributions are uncorrelated.  Immediately then, player $P_x$
optimizes their expected payoff via satisfying the condition
\begin{equation}
 \frac{d \langle \Pi_x \rangle}{d [P_{1x}(1)]}  = 1.
\end{equation}
Consequently, player $P_x$ optimizes their first and final stage
payoff by setting $P_{1x}(1)=1$ and so defects in this first
stage.  The shorthand notation $H_{n}=\{H_{n-1},y_n\}$ for $n\geq
1$ and some algebra allows writing the optimization conditions for
player $P_y$ as the set of simultaneous equations
\begin{eqnarray}
 \frac{d \langle \Pi_y \rangle}{d [P_{1y}(1)]}&=& \dots,  \nonumber \\
 &\vdots&   \nonumber \\
 \frac{d \langle \Pi_y \rangle}{d [P_{(N-1)y,H_{N-2}}(1)]} &=& -1 + \nonumber \\
  && \hspace{-3cm} + \sum_{y_1\dots y_{N-2}} P_{1y}(y_1) \dots
  P_{(N-2)y,H_{N-3}}(y_{N-2}) \times \nonumber \\
  && \hspace{-3cm} \sum_{y_N} y_N
  \left[ P_{Ny,\{H_{N-2},1\}}(y_N)-
  P_{Ny,\{H_{N-2},0\}}(y_N)\right],
 \nonumber \\
 \frac{d \langle \Pi_y \rangle}{d [P_{Ny,H_{N-1}}(1)]} &=& 1.
\end{eqnarray}
Hence, player $P_y$ optimizes their payoff by setting
$P_{Ny,H_{N-1}}(1)=1$ for every history set $H_{N-1}$, and by
setting $P_{(N-1)y,H_{N-2}}(1)=0$ for every history set $H_{N-2}$,
and eventually by setting $P_{ny,H_{n-1}}(1)=0$ for $1\leq n\leq
(N-1)$. These conditions locate the constrained equilibria at the
point $(x_1,y_1,\ldots,y_N)=(1,0,\dots,0,1)$ generating the play
sequence
\begin{equation}
 (x_n,y_n) = (D,C) (C,C) \dots (C,C) (C,D)
\end{equation}
to give expected payoffs
$(\langle\Pi_x\rangle,\langle\Pi_y\rangle) = (2N-1,2N-1)$. Here,
player $P_x$ defects in the first stage as their opponent cannot
respond without decreasing their payoff, while $P_y$ can defect in
the last stage when $P_x$ can no longer respond.

\begin{table*}[htb]
\begin{ruledtabular}
\begin{tabular}{c|cccccccc}
  $(\langle\Pi_x\rangle,\langle\Pi_y\rangle)$ & $j=0$     & $1$                               & $2$                                & $3$                               & $4$                                &  $\cdots$ & $N-2$                             & $N-1$                          \\ \hline
      $k=0$           & $\ddot{N},\ddot{N}$               & $\ddot{N}^{-2},\ddot{N}^{+1}$     &                 =                  &                =                  &                =                   &  $\cdots$ & $\ddot{N}^{-2},\ddot{N}^{+1/-2}$  & $\ddot{N}^{-1},\ddot{N}^{-1}$  \\
      $1$             & $\ddot{N}^{+1},\ddot{N}^{-2}$     & $\ddot{N}^{-1},\ddot{N}^{-1}$     & $\ddot{N}^{-3},\ddot{N}$           &                =                  &                =                   &  $\cdots$ & $\ddot{N}^{-3},\ddot{N}^{+0/-3}$  & $\ddot{N}^{-2},\ddot{N}^{-2}$  \\
      $2$             &       "                           & $\ddot{N},\ddot{N}^{-3}$          & $\ddot{N}^{-2},\ddot{N}^{-2}$      & $\ddot{N}^{-4},\ddot{N}^{-1}$     &                =                   &  $\cdots$ & $\ddot{N}^{-4},\ddot{N}^{-1/-4}$  & $\ddot{N}^{-3},\ddot{N}^{-3}$  \\
      $3$             &       "                           &       "                           & $\ddot{N}^{-1},\ddot{N}^{-4}$      & $\ddot{N}^{-3},\ddot{N}^{-3}$     &  $\ddot{N}^{-5},\ddot{N}^{-2}$     &  $\cdots$ & $\ddot{N}^{-5},\ddot{N}^{-2/-5}$  & $\ddot{N}^{-4},\ddot{N}^{-4}$  \\
      $4$             &       "                           &       "                           &             "                      & $\ddot{N}^{-2},\ddot{N}^{-5}$     &  $\ddot{N}^{-4},\ddot{N}^{-4}$     &  $\cdots$ & $\ddot{N}^{-6},\ddot{N}^{-3/-6}$  & $\ddot{N}^{-5},\ddot{N}^{-5}$  \\
      \vdots          & \vdots                            & \vdots                            & \vdots                             &  \vdots                           &  \vdots                            &  $\cdots$ & \vdots                            & \vdots                         \\
      $N-4$           &       "                           &       "                           &             "                      &            "                      &  $\ddot{N}^{-3},\ddot{N}^{-6}$     &  $\cdots$ & $\dot{N}^{+2},\dot{N}^{+5/+2}$    & $\dot{N}^{+3},\dot{N}^{+3}$    \\
      $N-3$           & $\ddot{N}^{+1/-2},\ddot{N}^{-2}$  & $\ddot{N}^{+0/-3},\ddot{N}^{-3}$  & $\ddot{N}^{-1/-4},\ddot{N}^{-4}$   & $\ddot{N}^{-2/-5},\ddot{N}^{-5}$  &  $\ddot{N}^{-3/-6},\ddot{N}^{-6}$  &  $\cdots$ & $\dot{N}^{+1},\dot{N}^{+4/+1}$    & $\dot{N}^{+2},\dot{N}^{+2}$    \\
      $N-2$           &                      "            &              "                    &            "                       &            "                      &             "                      &  $\cdots$ & $\dot{N}^{+2},\dot{N}^{+2}$       & $\dot{N}^{+1},\dot{N}^{+1}$    \\
      $N-1$           & $\ddot{N}^{-1},\ddot{N}^{-1}$     & $\ddot{N}^{-2},\ddot{N}^{-2}$     & $\ddot{N}^{-3},\ddot{N}^{-3}$      & $\ddot{N}^{-4},\ddot{N}^{-4}$     &  $\ddot{N}^{-5},\ddot{N}^{-5}$     &  $\cdots$ & $\dot{N}^{+1},\dot{N}^{+1}$       & $\dot{N},\dot{N}$.             \\
\end{tabular}
\end{ruledtabular}
\caption{\label{tab_equil} A partial listing of constrained
equilibria for the IPD when player $P_x$ implements a Markovian
strategy algorithm MKV-$k$I, while player $P_y$ implements an
MKV-$j$I algorithm.  Here, every shown payoff pair is a
constrained equilibrium point making selection of a single best
payoff maximization strategy difficult. For brevity, we restrict
consideration to large $N>8$ say, and write $\dot{N}^{\pm k}=(N\pm
k)$ and $\ddot{N}^{\pm k}=(2N\pm k)$. Fractional indices ($+k/-j$)
indicate alternate equilibria with payoff increments of $+k$ and
$-j$ respectively. Ditto signs (") and equal signs (=) copy values
downwards and to the right respectively. }
\end{table*}

Next we treat another combination of constraints assuming the
Markovian strategy algorithms for both players as
\begin{eqnarray}     \label{eq_TFT_xy}
  x_n & = & y_{n-1}  \nonumber \\
  y_n & = & x_{n-1}  ,
\end{eqnarray}
for $2\leq n\leq N$, equivalent to the conditioned probability
distributions $P_{nx}(x_n|H_{n-1})=\delta_{x_n,y_{n-1}}$ and
$P_{ny}(y_n|H_{n-1})=\delta_{y_n,x_{n-1}}$. These constraint sets
project the most general expected payoff functions to
\begin{equation}
\langle \Pi_z \rangle =
   \sum_{x_1,y_1} P_{1x}(x_1) P_{1y}(y_1)\Pi_z(x_1,y_1)
\end{equation}
for $z\in\{x,y\}$, where for a given play sequence $(x_1,y_1)$,
the payoffs are
\begin{eqnarray}
  \Pi_x & = &  \left\{
   \begin{array}{ll}
 2N - \frac{N}{2}x_1 - \frac{N}{2} y_1, & N \mbox{ even}\\
      &    \\
 2N - \frac{N-3}{2}x_1 - \frac{N+3}{2}y_1, & N \mbox{ odd}\\
   \end{array}
          \right.    \nonumber \\
          \nonumber   \\
  \Pi_y & = &  \left\{
   \begin{array}{lc}
 2N - \frac{N}{2} x_1 - \frac{N}{2} y_1, & N \mbox{ even}\\
      &    \\
 2N - \frac{N+3}{2}x_1 - \frac{N-3}{2}y_1, & N \mbox{ odd}.\\
   \end{array}
          \right.
\end{eqnarray}
The $N$ stage supergame has now been exactly reduced to a single
stage game with variables $x_1$ and $y_1$ with payoff matrices,
for $N$ even of
\begin{equation}
  \begin{array}{cc}
      & P_y \\
    P_x &
    \begin{array}{c|cc}
      (\Pi_x,\Pi_y) & C                           &  D                          \\   \hline
            C       & (2N,2N)                     & (\frac{3}{2}N,\frac{3}{2}N) \\
            D       & (\frac{3}{2}N,\frac{3}{2}N) & (N,N),                      \\
    \end{array}
  \end{array}
\end{equation}
and for odd $N$ of
\begin{equation}
  \begin{array}{cc}
      & P_y \\
    P_x &
    \begin{array}{c|cc}
      (\Pi_x,\Pi_y) & C                    &  D                   \\   \hline
            C       & (2N,2N)              & \frac{3}{2}[N-1,N+1] \\
            D       & \frac{3}{2}[N+1,N-1] & (N,N).               \\
    \end{array}
  \end{array}
\end{equation}
Here, the constraining strategy functions have changed the
off-diagonal elements of the effective payoff matrix to modify
equilibria. As usual, the generated constrained equilibria are now
calculated by applying the assumption of bounded rationality so
all remaining distributions are uncorrelated. The expected payoff
functions are then
\begin{eqnarray}
  \langle\Pi_x\rangle & = &  \left\{
   \begin{array}{ll}
 2N - \frac{N}{2}P_{1x}(1) - \frac{N}{2} P_{1y}(1), & N \mbox{ even}\\
      &    \\
 2N - \frac{N-3}{2}P_{1x}(1) - \frac{N+3}{2} P_{1y}(1), & N \mbox{ odd}\\
   \end{array}
          \right.    \nonumber \\
          \nonumber   \\
  \langle\Pi_y\rangle & = &  \left\{
   \begin{array}{lc}
 2N - \frac{N}{2} P_{1x}(1) - \frac{N}{2} P_{1y}(1), & N \mbox{ even}\\
      &    \\
 2N - \frac{N+3}{2}P_{1x}(1) - \frac{N-3}{2}P_{1y}(1), & N \mbox{ odd}.\\
   \end{array}
          \right.
\end{eqnarray}
As usual, the constrained equilibria are located via
\begin{eqnarray} \label{dx1_dy1}
  \frac{\partial \langle\Pi_x\rangle}{\partial [P_{1x}(1)]} & = &  \left\{
   \begin{array}{ll}
 - \frac{N}{2}, & N \mbox{ even}\\
      &    \\
 - \frac{1}{2}(N-3), & N \mbox{ odd}\\
   \end{array}
          \right.    \nonumber \\
          \nonumber   \\
 \frac{\partial \langle\Pi_y\rangle}{\partial [P_{1y}(1)]} & = &  \left\{
   \begin{array}{lc}
 - \frac{N}{2}, & N \mbox{ even}\\
      &    \\
  - \frac{1}{2}(N-3), & N \mbox{ odd}.\\
   \end{array}
          \right.
\end{eqnarray}
These conditions select the equilibrium points $P_{1x}(1)=0$ and
$P_{1y}(1)=0$ or $(x_1,y_1)=(0,0)\equiv(C,C)$ for either $N$ even
or for $N$ odd and greater than 3, while for $N=1$ the equilibria
is $P_{1x}(1)=1$ and $P_{1y}(1)=1$ or
$(x_1,y_1)=(1,1)\equiv(D,D)$.  When $N=3$ these conditions are
satisfied for any values of $(x_1,y_1)$ requiring examination of
actual payoffs motivating the selection
$(x_1,y_1)=(0,0)\equiv(C,C)$. The generated sequences of play are
\begin{equation} \label{sequence}
\begin{array}{c|c|l|c}
 N       &  (x_1,y_1) &               & (\langle\Pi_x\rangle,\langle\Pi_y\rangle) \\  \hline
 1       &  (1,1)     & (DD)          & (1,1)  \\
 N\geq 2 &  (0,0)     & (CC)\dots(CC) & (2N,2N).
\end{array}
\end{equation}

The shear number of possible strategy functions which might be
adopted make it necessary to consider more general functional
classes.  To this end, we consider that each player adopts a
mixed Markovian-Independent strategy, denoted MKV-$k$I,
where the MKV strategy is chosen from the first to $(N-k)$-th
stages while $k$I indicates that IND strategies are adopted for
the last $k$ stages. For player $P_x$ then, an MKV-$k$I strategy
sets
\begin{equation}
  x_n = \left\{
   \begin{array}{ll}
     y_{n-1} & 2\leq n \leq N-k \\
      &  \\
     x_1,x_{N-k+1},\dots, x_N & \mbox{independent}.
   \end{array}
  \right.
\end{equation}
Similar strategy functions denoted MKV-$j$I are implemented by
$P_y$.  These constraints are equivalent to the correlated
probability distributions
$P_{nx}(x_n|H_{n-1})=\delta_{x_n,y_{n-1}}$ for $2\leq n \leq N-k$
and $P_{ny}(y_n|H_{n-1})=\delta_{y_n,x_{n-1}}$ for $2\leq n \leq
N-j$.

The general MKV-$k$I strategy subsumes a number of other strategy
functions of interest.  For instance, setting $k=N-1$ or $k=N$
makes all variables independent (IND), so
MKV-$(N-1)$I$=$MKV-$N$I$=$IND. More interestingly, this strategies
subsumes certain deterministic strategies. To see this, suppose
that players consider a deterministic strategy choice $D$ in the
last stage, and so implement the strategy MKV-$1$D. More
generally, players may also consider strategies MKV-$k$D forcing
the choice $D$ in the last $k$ stages. However, it is not
difficult to see that this class of deterministic strategies is
weakly dominated by the class of MKV-$k$I. For the same $k$,
MKV-$k$I guarantees an equal or larger payoff than achievable
using MKV-$k$D against any strategy algorithm of the opponent. In
particular, the motivation to defect at the last stage for a
larger payoff is taken into account in the strategy class
MKV-$k$I. Exactly similar considerations establish that MKV-$k$I
strategies weakly dominate Tit-For-Tat strategies which specify
cooperation in the first stage.

Although the class of strategies MKV-$k$I is small in comparison
to the set of all possible strategies, it contains enough
complexity to demonstrate novel equilibria in the IPD. We consider
player $P_x$ to implement strategy function MKV-$k$I, while player
$P_y$ implements MKV-$j$I, so the most general expected payoff
functions become
\begin{eqnarray}
\langle \Pi_z \rangle &=&
   \sum_{\stackrel{x_1,x_{N-k+1},\dots,x_N}{y_1,y_{N-j+1},\dots y_N}}
 P_{1x}(x_1) P_{1y}(y_1) \times \dots \nonumber \\
 &&   \dots \times P_{Ny,H_{N-1}}(y_N) \Pi_z
\end{eqnarray}
for $z\in\{x,y\}$, where the payoffs for a given play sequence
$(x_1,x_{N-k+1},\dots,x_N,y_1,y_{N-j+1}\dots,y_N)$ are
\begin{equation}
 \Pi_z =  2N + \sum_{n=1}^{N} A_{zn} x_n + \sum_{n=1}^{N} B_{zn} y_n
\end{equation}
with variables $A_{zn}$ and $B_{zn}$ as specified in Appendix
\ref{app_calculations}. The assumption of bounded rationality and
that all remaining distributions are uncorrelated now allows
calculating the respective constrained equilibria with the
optimized payoffs as shown in Table \ref{tab_equil} for all
combinations of $k$ and $j$.

Every listed payoff pair in Table \ref{tab_equil} is an actual
constrained equilibrium point optimizing payoffs given imposed
constraints. As noted previously, there is no generally accepted
method to choose between alternate Nash equilibria. However,
strategy algorithms are independently selectable by each player,
so we can think that these strategy algorithms and constrained
equilibria create a new game defined by Table \ref{tab_equil}. In
this game on the constrained equilibrium space, each strategy
algorithm becomes a strategy choice, and each equilibrium point
becomes a pair of payoffs.  In the case where each pair of
strategy algorithms defines unique equilibrium payoffs, it is
obvious that this table can be considered as a game matrix. Hence
standard Nash techniques can be applied to determine global
equilibria among the located constrained equilibria. However, we
note that in general we have to take care to deal with multiple
equilibria generated by a pair of strategy algorithms. By applying
the Nash equilibrium definition to Table \ref{tab_equil}, we
obtain global equilibria at $(k,j)$ for either $k=0$ and $3\leq
j\leq (N-2)$, or $j=0$ and $3\leq k\leq (N-2)$.

These global equilibria can be considered rational for the IPD in
this restricted class of strategies, and there is no established
way to select a particular one among these.  The more important
feature given from this analysis is that cooperation naturally
arises from these equilibria.  The pathways produced by these
equilibria are dominated by cooperation apart from some different
choices at the last stage. This cooperative behaviour is
caused by the imposition of strategy function constraints.

\section{Correlated play and the last stage of the IPD}
\label{sect_BI}

In the previous sections, we have emphasized the importance of
correlated play in supergame analysis and the assumption of
independent mixed or behavioural strategy choice variables
required by the Nash equilibrium definition. Although experiments
on the IPD (and other games) have shown significantly different
behaviours from that predicted by the Nash analysis, there has
been little motivation to revise this assumption and extend the
Nash definition of equilibrium to a fully correlated analysis.
Many alternate approaches have been proposed, and the field has
developed in the direction of explaining why people do not behave
as rationally as they could. In the IPD this explanatory emphasis
resulted largely because of the typical use of the backwards
induction (BI) argument. However, the results obtained by our
general correlated analysis differ from the predictions of the
typical BI argument, implying misuse of the BI optimizing
technique in application to the IPD analysis. In this section we
clarify the confusion introduced by the typical BI argument in the
IPD analysis.

As noted, BI is based on dynamic programming techniques, and like
any variational optimization technique, it is applied only to
uncorrelated independent variables. That is, any correlations or
constraints must be resolved before optimization commences.
Necessarily then, BI must derive the same solution as the Nash
analysis under the same assumed constraint set.  As typically
used, the BI argument in the IPD is commonly used to justify the
assumption of independent variables underlying the Nash
equilibrium solution. As is well known, the rationale commences
with the last stage of a finite supergame. In the typical BI
argument, the last stage has a special role, and the rest of the
argument follows in exactly the way by iteration. In particular,
the usual claim is that BI requires that the last stage of an IPD
is to be considered an independent stage. It is argued that,
although there are many reasons for players to cooperate, such as
the presence of any of long range considerations
\cite{Selten_78_12}, reputation effects \cite{Bermudez_99_24},
off-equilibrium pathway signals \cite{Pettit_99_17}, or
punishments and rewards \cite{Hargreaves_95}, at the last stage
none of these are present. Hence, it is claimed, the last stage is
an independent stage, i.e. a single PD. Once we have accepted the
last stage as a single PD, then under CKR, BI automatically
derives the unique Nash path. In this sense, BI is a complement of
the Nash analysis to establish the Nash unique solution of the
IPD.  This argument would succeed with the addition of a proof
that these, and only these effects permitted correlated play.
Unfortunately, this is not the case.

The significant difference introduced by allowing rational players
to adopt correlated play whenever that turns out to be payoff
maximizing, is that they will consider the IPD as a whole, which
they are able to do as unbounded rational agents. The correlated
analysis in the previous sections has shown that the first step of
the BI argument about the last stage is not necessarily optimal.
In addition, CKR by itself specifies nothing about whether the
last stage of an IPD is or is not equivalent to a single stage
game. However, under CKR, players of unbounded rationality must
take account of the payoffs available under correlated play. In
general then, the typical usage of a BI argument to justify a Nash
pathway as being optimal is not correct under CKR and correlated
strategies are required for an optimal solution. For the typical
BI argument to justify the Nash pathway as uniquely optimal, it is
necessary that the game analysis, and hence CKR, is restricted to
a particular kind of correlation, namely the assumption of
independent variables. Even though there are apparently no overt
motivations for players to cooperate in the last stage, it is not
optimal and hence not rational for players to even consider ``what
should we do if we are at the last stage?".

\section{Conclusion}

This paper defines novel constrained equilibria in the middle
ground between current definitions which require that supergame
strategy choices be either all independent or all fully specified
by strategy functions. We employ standard conditioned history
expansions of the joint correlated probability distributions
describing expected payoff functions which specify only some (or
no) strategy choice variables (a dependent set) in terms of other
variables (an independent set). We then apply standard
optimization procedures such as Nash equilibria procedures or
backwards induction to the composite payoff functions defined over
the remaining independent variables to locate novel constrained
equilibria. The methods developed in this paper ensure that there
is no conflict between game theoretic optimization techniques and
more general variational optimization procedures. We derive novel
constrained equilibria in the finite iterated prisoner's dilemma
showing in particular, that backwards induction establishes that
it can be payoff maximizing to cooperate in the finite IPD
including in the last stage of this game.  These results contrast
with existing claims that payoff maximization requires defection
in this last stage, but these results all depend on the {\it a
priori} assumption that choice variables are uncorrelated.  We
conclude by discussing the validity of typical arguments proving
that ALL DEFECT is the privileged optimal pathway in the IPD.

This paper derived novel constrained equilibria using general
Markovian-like strategy functions, and a broader analysis of the
many possible unknown contingent strategies requires a functional
representation for algorithms. A mathematical functional analysis
would allow us to consider multiple algorithms using a functional
metric to measure the ability of different algorithms to determine
supergame outcomes. However, the utility of such a functional
analysis remains an open question and will be dealt with in later
work.

Game theory was originally proposed to provide a simple analytic
environment for economic and social interactions and the analysis
of this paper has broader application to this wider sphere. For
instance, constrained equilibria eliminates the first-mover
advantage in iterated bargaining games and explains experimental
observations of more equitable play. In such applications however,
the lack of a systematic way to select a particular equilibrium
among others becomes a more serious and important problem. This
issue and the broader application to economics, social games and
evolutionary games will be addressed in later work.

\section{Acknowledgements}

MJG thanks Rodney Beard for helpful reviewing comments and for
partially translating Selten's paper in German modelling an
oligopoly with sticky demand \cite{Selten_65_30}.  KN acknowledges
the hospitality of Hewlett-Packard Laboratories in Bristol.

\appendix

\begin{widetext}

\section{Constrained equilibria in the IPD}
\label{app_calculations}

Suppose player $P_x$ adopts an MKV-$k$I strategy and player $P_y$
adopts an MKV-$j$I strategy function leaving independent variables
$(x_1,x_{N-k+1},\dots,x_{N})$ and $(y_1,y_{N-j+1},\dots,y_{N})$,
where we also assume $N\geq 3$.

For $1\leq k\leq (N-1)$ and $j=0$, the independent variables are
$(x_1,x_{N-k+1},\dots,x_{N})$ and $y_1$, and payoffs are
\begin{eqnarray}
  \Pi_x & = &  \left\{
   \begin{array}{lc}
 2N  + \frac{k-N}{2} x_1 + \frac{k-4-N}{2}y_1
          - \sum_{n=N-k+1}^{N-1} x_n + x_N, & (N-k) \mbox{ even}\\
      &    \\
 2N  + \frac{k-1-N}{2} x_1 + \frac{k-3-N}{2}y_1
          - \sum_{n=N-k+1}^{N-1} x_n + x_N, & (N-k) \mbox{ odd}.\\
   \end{array}
          \right.
          \nonumber   \\
  \Pi_y & = &  \left\{
   \begin{array}{ll}
 2N + \frac{k-N}{2} x_1 + \frac{2+k-N}{2}y_1
          - \sum_{n=N-k+1}^{N-1} x_n - 2 x_N, & (N-k) \mbox{ even}\\
      &    \\
 2N + \frac{k-1-N}{2} x_1 + \frac{3+k-N}{2}y_1
          - \sum_{n=N-k+1}^{N-1} x_n - 2 x_N, & (N-k) \mbox{ odd}.\\
   \end{array}
          \right.
\end{eqnarray}

For $1\leq k\leq (N-1)$ and $j=(N-1)$, the independent variables
are $(x_1,x_{N-k+1},\dots,x_{N})$ and $(y_1,\dots,y_N)$, and
payoffs are
\begin{eqnarray}
  \Pi_x & = &  2N + x_1 + \sum_{n=N-k+1}^{N} x_n
        - \sum_{n=1}^{N-k-1} y_n - 2 \sum_{n=N-k}^{N} y_n , \nonumber \\
  \Pi_y & = &  2N - 2 x_1 - 2 \sum_{n=N-k+1}^{N} x_n
        - \sum_{n=1}^{N-k-1} y_n + \sum_{n=N-k}^{N} y_n.
\end{eqnarray}

For $1\leq k=j\leq (N-1)$, the independent variables are
$(x_1,x_{N-k+1},\dots,x_{N})$ and $(y_1,y_{N-k+1},\dots,y_{N})$,
and payoffs are
\begin{eqnarray}
  \Pi_x & = &  \left\{
   \begin{array}{ll}
 2N + \frac{k-N}{2}x_1+ \frac{k-N}{2} y_1
          + \sum_{n=N-k+1}^{N} x_n - 2 \sum_{n=N-k+1}^{N} y_n, & (N-k) \mbox{ even}\\
      &    \\
 2N + \frac{3+k-N}{2}x_1+ \frac{k-3-N}{2} y_1
          + \sum_{n=N-k+1}^{N} x_n - 2 \sum_{n=N-k+1}^{N} y_n, & (N-k) \mbox{ odd}\\
   \end{array}
          \right.    \nonumber \\
          \nonumber   \\
  \Pi_y & = &  \left\{
   \begin{array}{lc}
 2N + \frac{k-N}{2}x_1+ \frac{k-N}{2} y_1
          - 2 \sum_{n=N-k+1}^{N} x_n + \sum_{n=N-k+1}^{N} y_n, & (N-k) \mbox{ even}\\
      &    \\
 2N + \frac{k-3-N}{2}x_1+ \frac{3+k-N}{2} y_1
          - 2 \sum_{n=N-k+1}^{N} x_n + \sum_{n=N-k+1}^{N} y_n, & (N-k) \mbox{ odd}.\\
   \end{array}
          \right.
\end{eqnarray}

For $1\leq k\leq (N-1)$ and $1\leq j\leq (N-1)$ with $k>j$, the
independent variables are $(x_1,x_{N-k+1},\dots,x_{N})$ and
$(y_1,y_{N-j+1},\dots,y_{N})$, and payoffs are
\begin{eqnarray}
  \Pi_x & = &  \left\{
   \begin{array}{ll}
 2N + \frac{k-N}{2}x_1+ \frac{k-4-N}{2} y_1
          - \sum_{n=N-k+1}^{N-j-1} x_n
          + \sum_{n=N-j}^{N} x_n
          - 2 \sum_{n=N-j+1}^{N} y_n, & (N-k) \mbox{ even}\\
      &    \\
 2N + \frac{k-1-N}{2}x_1+ \frac{k-3-N}{2} y_1
          - \sum_{n=N-k+1}^{N-j-1} x_n
          + \sum_{n=N-j}^{N} x_n
          - 2 \sum_{n=N-j+1}^{N} y_n, & (N-k) \mbox{ odd}\\
   \end{array}
          \right.    \nonumber \\
          \nonumber   \\
  \Pi_y & = &  \left\{
   \begin{array}{lc}
 2N + \frac{k-N}{2}x_1+ \frac{2+k-N}{2} y_1
          - \sum_{n=N-k+1}^{N-j-1} x_n
          - 2 \sum_{n=N-j}^{N} x_n
          + \sum_{n=N-j+1}^{N} y_n, & (N-k) \mbox{ even}\\
      &    \\
 2N + \frac{k-1-N}{2}x_1+ \frac{3+k-N}{2} y_1
          - \sum_{n=N-k+1}^{N-j-1} x_n
          - 2 \sum_{n=N-j}^{N} x_n
          + \sum_{n=N-j+1}^{N} y_n, & (N-k) \mbox{ odd}.\\
   \end{array}
          \right.
\end{eqnarray}
\end{widetext}


\begin{thebibliography}{10}
\bibitem{Correlations}
{\footnotesize The correlation between variables $z$ and $z'$ is
  \begin{displaymath} {\rm cor}(z,z') = \frac{{\rm
  cov}(z,z')}{\sigma_z\sigma_{z'}}, \nonumber \end{displaymath} defined in
  terms of the covariance \begin{displaymath} {\rm cov}(z,z') = \langle z
  z'\rangle - \langle z\rangle \langle z'\rangle, \nonumber \end{displaymath}
  the variance $\sigma_z^2={\rm cov}(z,z)$, and mean $\langle z\rangle$
  \protect\cite{Kelly_94}}.
\bibitem{Information}
{\footnotesize The mutual information between two discrete
probabilistic
  variables $x$ and $y$ is $H(x;y)=H(x)-H(x|y)$ written in terms of the entropy
  (or uncertainty) of $x$, $H(x)=-\sum_x P(x) \log P(x)$, and the conditional
  entropy of $x$ given $y$, $H(x|y)=-\sum_{x,y}P(x,y)\log P(x|y)$, while the
  joint entropy or uncertainty of $x$ and $y$ is $H(x,y)=-\sum_{x,y}P(x,y)\log
  P(x,y)$. Then, when variables $x$ and $y$ are uncorrelated, their mutual
  information is minimized at $H(x;y)=0$ while their joint entropy or
  uncertainty is maximized at $H(x,y)=H(x)+H(y)$. Conversely, when these
  variables are perfectly correlated, their mutual information is maximized at
  $H(x;y)=H(x)$ while their joint entropy or uncertainty is minimized at
  $H(x,y)=H(x)$ \protect\cite{Cover_91}}.
\end{thebibliography}

\end{document}